\documentclass[12pt]{article}   
\newlength{\dinwidth}
\newlength{\dinmargin}
\def\NIM#1#2#3 {{\em Nucl. Instrum. Methods} {\bf#1} (#2) #3}

\def\pl#1#2#3   {{\em Phys. Lett.} {\bf#1} (#2) #3}
\def\np#1#2#3   {{\em Nucl. Phys.} {\bf#1} (#2) #3}
\def\prev#1#2#3 {{\em Phys. Rev.} {\bf#1} (#2) #3}
\def\cpc#1#2#3  {{\em Computer Phys. Comm.} {\bf#1} (#2) #3}
\def\eurj#1#2#3 {{\em Eur. Phys. J.} {\bf#1} (#2) #3}
\def\lsim{\mathrel{\rlap{\lower4pt\hbox{\hskip1pt$\sim$}}
    \raise1pt\hbox{$<$}}}                % less than or approx. symbol
\def\gsim{\mathrel{\rlap{\lower4pt\hbox{\hskip1pt$\sim$}}
    \raise1pt\hbox{$>$}}}                % greater than or approx. symbol
\usepackage{glas,epsfig} 
\begin{document}
\begin{titlepage}{GLAS-PPE/1999-07}{July 1999}
\title{Comparison of Dijet Cross Sections vs Q$^2$ with \\
Various Monte Carlo Models}

\author{N. Macdonald$^a$}
\centerline{$^a$ Dept. of Physics and Astronomy,}
\centerline{University of Glasgow, Glasgow, U.K.}
\begin{abstract}
The dijet cross section as a function of both the fraction of the photon
momentum participating in the hard process, $x_\gamma^{OBS}$, and the 
photon's virtuality, Q$^2$, 
is compared with the predictions of Herwig 5.9 for 
various photon parton distribution functions.
The ratio of dijet cross sections with $x_\gamma^{OBS} < 0.75$ and
$x_\gamma^{OBS} > 0.75$ is measured as a function of $Q^2$.
This ratio is found to decrease as $Q^2$ increases consistent with the
hypothesis that the photon parton distribution functions decrease with
increasing photon virtuality.
\end{abstract}
\end{titlepage}

\section{Introduction}

Experimental information on the partonic structure of the photon is
obtained from the HERA ep
collider experiments via measurements of jet photoproduction~\cite{prevpap}. 
Leading order (LO) QCD predicts that photon interactions
have a two-component nature.
In direct photon processes the entire momentum of the photon takes part in
the hard subprocess with a parton from the proton 
whereas in resolved photon processes, the photon acts as a source of
partons and one of these enters the hard subprocess.
By tagging inclusive dijet events (two or more jets) in conjunction
with tagging the scattered electron, information on
the structure of the photon can be extracted as a function of the
virtuality of the photon, Q$^2$. 

\section{Direct and Resolved Photoproduction}

Jet photoproduction events are split into two classes

\begin{enumerate}
\item Direct photoproduction. Here the virtual photon
connects directly to a quark line in the Feynman diagram, the quark
absorbing the photon. The fraction of the photon's four
momentum entering the hard dijet subprocess, $x_\gamma^{LO}$, 
is equal to unity. 
\item Resolved photoproduction. Here the virtual photon
behaves as a source of partons, one of which takes part in the hard
subprocess. One of the main 
signatures of a resolved process is that of the low $p_T$ photon
remnant, the partons produced by the photon which did not 
take part in the scatter. The fraction of the photon's four 
momentum entering the hard dijet subprocess, $x_\gamma^{LO}$,
 is now less than unity. 
\end{enumerate}
An example of each process is shown in figure~1. On the
left is an example of the diagram corresponding to a direct process,
on the right is an example of a resolved process. 

\begin{figure}[htb]
\centering
\makebox[5.8cm]{\epsfig{file=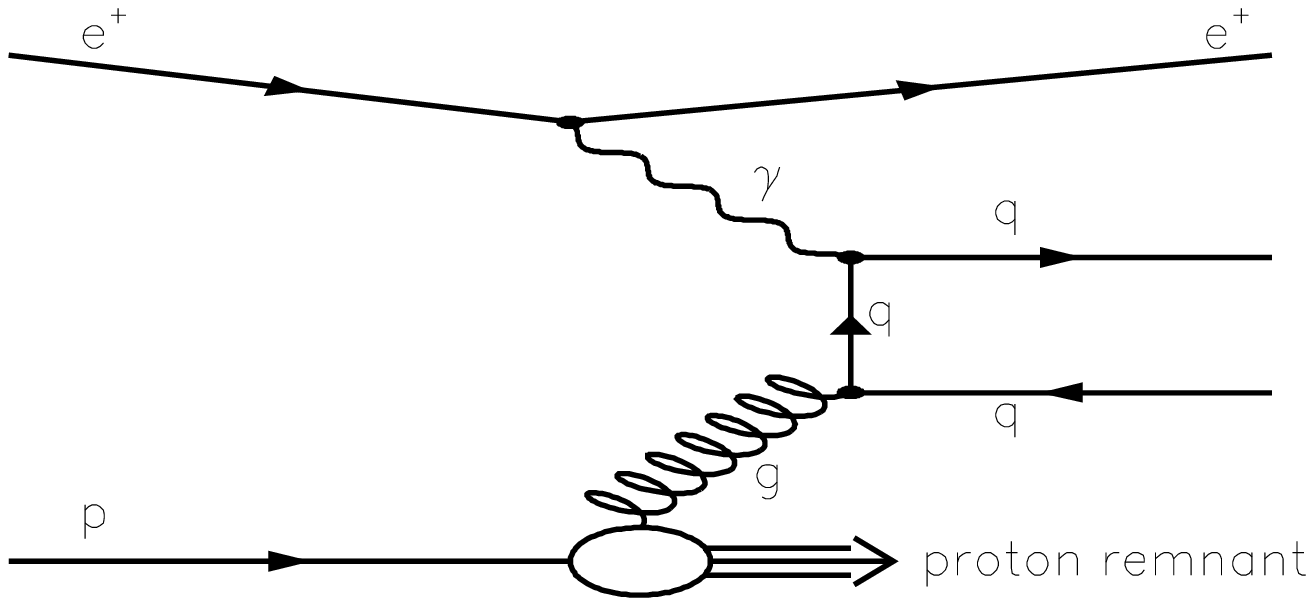,width=5.0cm,height=3.5cm,angle=0}}
\makebox[5.8cm]{\epsfig{file=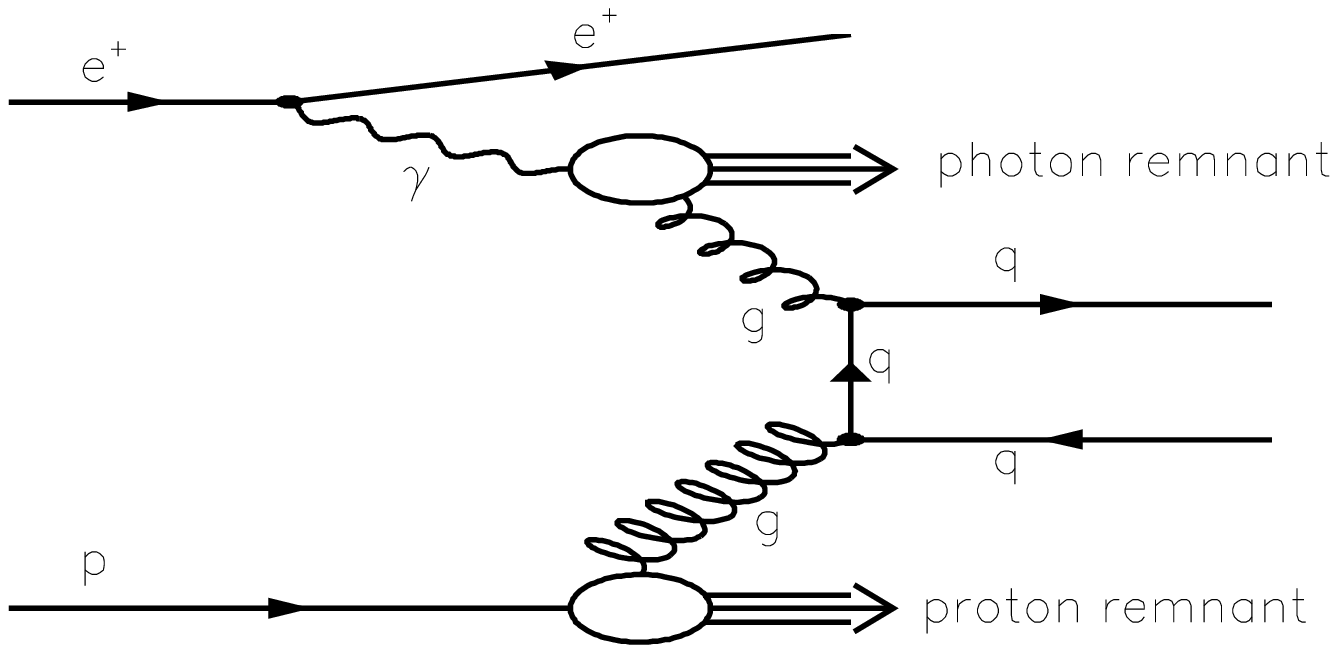,width=5.0cm,height=3.5cm,angle=0}}
\label{Fdiags}
\caption{\it Feynman diagrams of direct and resolved process}
\end{figure}

\subsection{$\bf x_\gamma^{obs}$}

Experimentally $x_\gamma^{LO}$ cannot be measured directly, so 
the variable $x_\gamma^{obs}$ is chosen as an estimator of the
fraction of the photon's momentum that takes part in the hard
scatter. $x_\gamma^{obs}$ 
is then the fraction of the photon momentum manifest in the two
highest $p_T$ jets and is defined by the equation
\begin{eqnarray}
x_\gamma^{obs} = \frac{\sum_{j=1}^2 E_{Tj} e^{- \eta_j}}{2E_e y} =
\frac{\sum_{j=1}^2 (E_j - p_{z_j})}{\sum_{hadrons} (E - p_z)} \nonumber
\end{eqnarray}
where $E_{Tj}$ is the transverse energy of jet $j$, $\eta_j$ is the
pseudorapidity of the jet, and $y$ is the inelasticity of the event.
\par
The fact that jets of hadrons are measured, 
and not partons means that the value of $x_\gamma^{obs}$ is
dependent upon the jet finding algorithm used, and hadronisation
effects. 
Events with a high value of $x_\gamma^{obs}$ are mostly direct
processes, low $x_\gamma^{obs}$ resolved processes. 
As a result of this, ``direct enriched'' events are classed 
as being those with
$x_\gamma^{obs} > 0.75$, and ``resolved enriched'' events as those with
$x_\gamma^{obs} < 0.75$. This is an experimental definition, since
there are still true resolved events with $x_\gamma^{obs} > 0.75$,
and vice versa, but the contamination between the two event classes is
minimised by having the cut at this value.

\section{Comparison of Data Cross Sections with Monte Carlo}

The cross section that is measured is defined by the following set of
kinematic cuts 

\begin{itemize}
\item Inclusive dijets ($k_T$ clustering algorithm) \cite{kt}
\item $Q^{2} < 1.0$, $0.1 < Q^{2} < 0.55$, $1.5 < Q^{2} < 4.5$ (GeV$^2$)
\item $0.20 < y < 0.55$
\item $E_{T}^{jets} > 5.5$ GeV
\item $-1.125 < {\eta}^{jets} < 2.2$
\end{itemize}

Three different Q$^2$ regions are accessed, firstly Q$^{2} < 1.0$
GeV$^2$ corresponds to quasi-real photons, where the electron is not
tagged, and the large bulk of such events lead to a median Q$^2$ of
0.001 GeV$^2$. For $0.1 < Q^{2} < 0.55$ GeV$^2$ ZEUS has a small angle
electron detector \cite{BPCNIM,BPCF2} for tagging events in the
transition region between 
photoproduction and DIS. For $1.5 < Q^{2} < 4.5$ GeV$^2$ the electron
is tagged in the ZEUS main calorimeter \cite{CAL}. 
\par
\begin{figure}[htb]
\centering
\epsfig{file=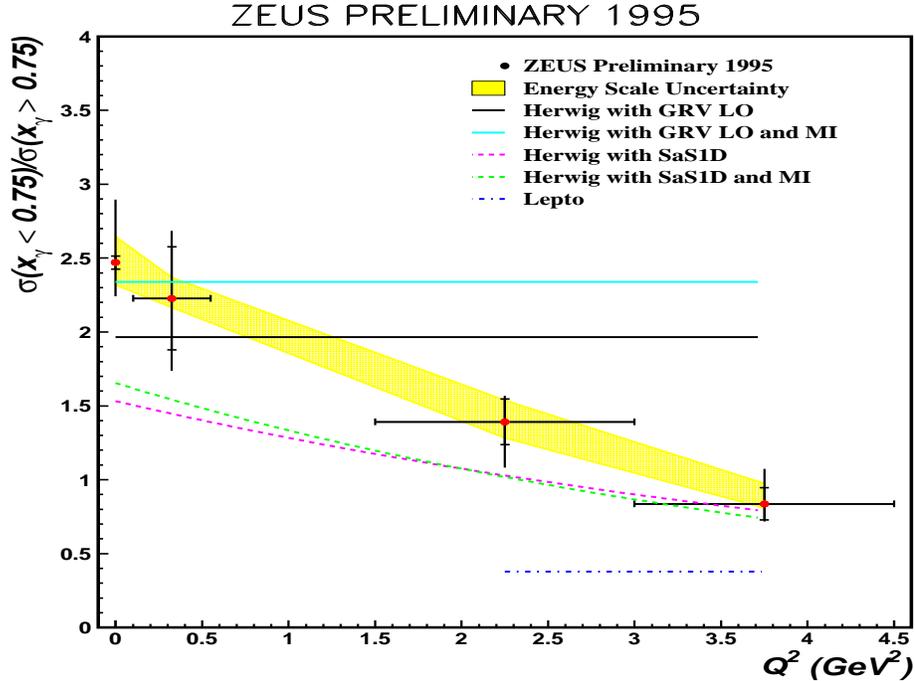,width=12.0cm,height=9.0cm,angle=0}
\label{ratio}
\caption{\it Ratio of resolved enriched to direct enriched dijet cross
sections. Theory curves are produced by Herwig 5.9, with various
photon PDFs. Also shown is the result of Lepto 6.5.1}
\end{figure}

Figure~\ref{ratio} contains the dijet cross sections shown as a ratio of 
resolved enriched over direct enriched as a function of Q$^2$. 
Since certain systematic errors cancel in the ratio, this is a more 
precise measurement than absolute differential cross sections. 
The ratio is also sensitive in shape (but not
normalisation) to the photon PDF, even when the absolute cross section given
by Herwig 5.9 \cite{HERWIG} does not match the data. One can see that
GRV LO \cite{GRV} is 
flat, as expected, since it has no
Q$^2$ dependence. The effect of multi-parton interactions is to
increase the ratio, but again for GRV LO this is flat vs Q$^2$. 
\par
The Lepto 6.5.1 \cite{lepto} prediction is also flat since 
this contains direct
processes only. Even with direct only processes, the ratio of resolved
enriched over direct enriched cross sections is non-zero,
demonstrating that $x_\gamma^{obs}$ is sensitive to the jet
finding algorithm and hadronisation effects, and hence has a reduced
correlation with $x_\gamma^{LO}$. The data is approaching this direct
only limit, however resolved processes
are still required in order to describe the data up to a Q$^2$ of
4.5~GeV$^2$. 
\par
The Schuler and Sj\"ostrand (SaS1D) photon PDF \cite{sas} does 
have a suppresion of parton densities in the
photon vs Q$^2$, and this can be clearly seen from the shape of the
curves. The effect of multi-parton interactions is small for this PDF,
and both curves approach the data for the higher Q$^2$ bins, whilst
disagreeing for lower Q$^2$ values. Given the size of the errors on
the data, the shape of these curves is consistent with the ZEUS
data.

% /data/hibernian/uk/macdonld/hztool/standard/kumacs/plotb.kumac
% produces this plot

\end{document}